\documentclass[letter,twocolumn]{jpsj3}
\usepackage{txfonts}
\usepackage{graphicx,bm,color,braket}

\title{Hall Effect in the Abrikosov Lattice of Type-II Superconductors}

\author{ Wataru Kohno, Hikaru Ueki, and Takafumi Kita}
\inst{Department of Physics, Hokkaido University, Sapporo 060-0810, Japan} 
\abst{
We study vortex charging caused by the Lorentz force on supercurrent based on the augmented quasiclassical equations of superconductivity. 
Our numerical study on an $s$-wave vortex lattice in the range $H_{\rm c1}<H<H_{\rm c2}$ reveals that each vortex core with a single flux quantum also accumulates charge 
 due to the circulating supercurrent and has a Hall voltage across the core.
The field dependence of the charge density at the core center is well described by $\rho({\bf{0}})\propto H(H_{{\rm{c}}2}-H)$
with a peak near $H_{\rm c2}/2$ originating from competition between the increasing magnetic field and
the decreasing pair potential. 
The peak value of the accumulated charge in a core region of radius $0.5\xi_0$ is estimated to be
about $\eta\Delta_{0}/(k_{\rm F}\xi_0)\times|e|$ C per $\Delta z=10$ $\AA$ along the flux line at low temperatures, 
where $\eta\equiv\pi\epsilon_0\Delta z/|e|^2=1.09\times10^{18}$ ${\rm{J^{-1}}}$ with $e<0$ the charge of an electron,  $\Delta_0$ the energy gap at $T=0$,
$k_{\rm F}$ the Fermi wave number, and $\xi_0$ the coherence length at $T=0$.}

\begin{document}
\maketitle

The Lorentz force on electric currents flowing in magnetic fields has a unique component perpendicular to both the current and field.
It generally induces charge redistribution before recovering a steady state to produce a Hall voltage that eventually brings about force balance 
along the transverse direction.
Extensive studies have been performed over the last few decades on this {\em Hall effect}\cite{Hall} in metals and semiconductors, especially on the quantum Hall effect in two dimensions.\cite{Girvin}

In contrast, 
we still have little understanding of the phenomena in superconductors.
This is because the force on supercurrent itself may easily be overlooked in the presence of the predominant diamagnetic effect by supercurrent
obeying  Amp\`ere's law.
Indeed, the Lorentz force is missing from the Ginzburg--Landau\cite{GL} and Eilenberger\cite{Eilenberger} equations
that have been used extensively in the literature,\cite{Parks69,Kopnin,KitaText}
and can only be reproduced microscopically as a next-to-leading-order contribution in the expansion of the Gor'kov equations
in terms of the quasiclassical parameter $\delta\equiv 1/k_{\rm F}\xi_0$.\cite{Kita01,Ueki}
Hence, the physics of the Lorentz force in superconductors remains mostly theoretically unexplored.

This Hall effect in superconductors may be divided into two categories: one in equilibrium with persistent currents\cite{Kita09,Ueki,VS64,AW68,Kel1,Kel2}
and the other in nonequilibrium situations with the motion of vortices and dissipation.\cite{KK76,Kopnin,AK,Hagen93,Nagaoka98}
The first one is inherent to superconductors and easier to handle but nevertheless has not been paid much attention in the literature.
We here focus on this first category and study vortex charging in type-II superconductors as a function of the magnetic field
based on the augmented quasiclassical equations of superconductivity.\cite{Kita01,Ueki}

It has been previously shown that charge accumulation due to the Lorentz force occurs in equilibrium near edges in the Meissner state\cite{Kita09,VS64,AW68,Kel1,Kel2}
and also around the core of an isolated vortex slightly above the lower critical field $H_{{\rm c1}}$.\cite{Ueki} 
With these results and noting that the Lorentz force is proportional to the flux density $B$, 
we expect an enhancement of the charging in finite magnetic fields.
To clarify this point, we here consider an $s$-wave vortex lattice and 
calculate its charge distribution numerically as a function of the magnetic field from $H_{\rm c1}$ up to the upper critical field $H_{\rm c2}$ at various temperatures.
It is thereby shown that the charge density at the vortex center has a strong field dependence 
with a peak around $H_{\rm c2}/2$ whose value is 10--100 times larger than that of the isolated vortex.
This field dependence is characteristic of charging by the Lorentz force making it experimentally distinguishable from other possible mechanisms 
for vortex-core charging.\cite{KF,Feigel'man,Blatter,Hayashi,MH,Chen,Knapp}
In this context, Khomskii and Freimuth\cite{KF} and Feigel'man and coworkers \cite{Feigel'man,Blatter}  proposed different mechanisms 
by regarding the core region as the normal state and 
considering its chemical-potential difference from the surroundings.\cite{KF,Feigel'man,Blatter} 
However, the resulting charge accumulation, if any, should 
decrease monotonically as $H$ is increased because of the decreasing pair potential.

We now study the charge and electric-field distributions due to the Lorentz force in the vortex lattice of a clean $s$-wave type-II superconductor.
To this end, we use the augmented quasiclassical equations in the Matsubara formalism.\cite{Ueki} As shown in Ref.\ \citen{Ueki},
they can be decoupled into an electric-field equation plus the standard Eilenberger equations 
through an expansion in terms of the dimensionless quasiclassical parameter $\delta\ll 1$.
The Eilenberger equations are reproduced in this procedure by collecting terms of $O(1)$ in the expansion.
They are given by\cite{Kopnin,KitaText,Rainer83,LO86}
\begin{subequations}
\label{Ei}
\begin{align}
&\omega_n f+\frac{1}{2}\hbar{\bm v}_{\rm{F}}\cdot \left(\bm{\nabla}-i\frac{2e{\bm A}}{\hbar}\right)f=\Delta g,\label{Ei1}\\
&\Delta=g_0 \pi  k_{\rm{B}} T \sum_{n=-\infty}^{\infty}\langle f \rangle_{\rm{F}}\label{Ei3},\\
&\bm{\nabla}\times\bm{\nabla}\times{\bm A}=-i2\pi e\mu_0N(0)k_{\rm{B}}T\sum_{n=-\infty}^{\infty}\langle {\bm v}_{\rm{F}}g \rangle_{\rm{F}}\label{Ei4},
\end{align}
\end{subequations}
with the normalization condition $g={\rm{sgn}}(\omega_n)\left(1-f\bar{f}\right)^{1/2}$.
Here $\omega_n\equiv (2n+1)\pi k_{\rm{B}}T$ $(n=0,\pm1,\cdots)$ with $k_{\rm{B}}$ and $T$ denoting the Boltzmann constant and temperature,
$f=f(\omega_n , {\bm p}_{\rm F}, {\bm r})$ is the anomalous quasiclassical Green's function and $\bar{f}\equiv f^{*}(\omega_n , -{\bm p}_{\rm F}, {\bm r})$,
${\bm p}_{\rm F}$ and ${\bm v}_{\rm F}$ are the Fermi momentum and Fermi velocity,
and $e<0$ is the charge of an electron, respectively. 
Equations (\ref{Ei3}) and (\ref{Ei4}) are the self-consistency equation for the pair potential 
${\Delta}({\bm r})$ and Maxwell's equation (Amp\`ere's law) for the vector potential
${\bm A}({\bm r})$, respectively, where $g_0\ll 1$ is a dimensionless coupling constant responsible for the Cooper pairing, $\langle\cdots\rangle_{\rm F}$ 
denotes the Fermi surface average normalized as $\langle 1\rangle_{\rm F}=1$,
$N(0)$ is the normal density of states per spin and unit volume at the Fermi energy, and $\mu_0$ is the vacuum permeability.

The Lorentz force emerges as a correction of $O(\delta)$, which is shown to induce an electric field ${\bm E}$ that obeys\cite{Kita09,Ueki}
\begin{align}
&(-\lambda^2_{\rm{TF}}\bm{\nabla}^2+1){{\bm E}}=-i\pi k_{\rm{B}}T{\bm B}\times\sum_{n=-\infty}^{\infty}\left\langle\frac{\partial g}{\partial{\bm p}_{\rm{F}}}  \right\rangle_{\rm{F}},\label{el}
\end{align}
where $\lambda_{\rm TF} \equiv \sqrt{\epsilon_0/2 e^2 N (0)}$ is the Thomas--Fermi screening length
with $\epsilon_0$ the vacuum permittivity. 
Note that the source term on the right-hand side of Eq.\  (\ref{el}) is given by the solutions $g$ and ${\bm B}={\bm\nabla}\times {\bm A}$ from Eq.\ (\ref{Ei}).
Hence, we can calculate electric fields generated by the Hall effect in various external circumstances based on the solutions of Eq.\ (\ref{Ei}).

Here, we use Eq.\ (\ref{Ei}) to construct 
vortex-lattice solutions of an $s$-wave pairing on a two-dimensional isotropic Fermi surface, which is perpendicular to the magnetic field.
The corresponding vector potential is expressible in terms of the average flux density $\bar{\bm B}=(0,0,\bar{B})$ 
as ${\bm A}({\bm r})=(\bar{\bm B}\times{\bm r})/2+\tilde{\bm A}({\bm r})$,\cite{KitaText,KitaGL}
where $\tilde{\bm A}$ describes the spatial variation of the flux density.
Functions $\tilde{\bm A}$ and $\Delta$ for the triangular lattice obey the following periodic boundary conditions:\cite{Klein,Ichioka1,KitaGL}
\begin{subequations}
\label{sym}
\begin{align}
&\tilde{\bm A}({\bm r}+{\bm R})=\tilde{\bm A}({\bm r})\label{sym1},\\
&\Delta({\bm r}+{\bm R})=\Delta({\bm r})\exp\left[i\frac{|e|}{\hbar}{\bm B}\cdot\left({\bm r}\times{\bm R}\right)+i\pi n_1n_2\right]\label{sym2},
\end{align}
\end{subequations}
where ${\bm R}=n_1{\bm a}_1+n_2{\bm a}_2$ with $n_1$ and $n_2$ denoting integers, and ${\bm a}_1=a_2(\sqrt{3}/2,1/2,0)$ and ${\bm a}_2=a_2(0,1,0)$ are 
the basic vectors of the triangular lattice with length $a_2$ determined by the flux-quantization condition $({\bm a}_1\times{\bm a}_2)\cdot\bar{\bm B}=h/2|e|$.

Equation (\ref{Ei}) with the boundary condition in Eq.\ (\ref{sym}) can be solved iteratively for a given set of $(T,\bar B)$.
It is convenient for this purpose to transform Eq.\ (\ref{Ei1}) into the Riccati form.\cite{NNH93,SM,KitaText}
In the first iteration, we substitute the trial functions
\begin{align}
\Delta({\bm r})=\Delta_{T}\sqrt{1-\frac{\bar{B}}{B_{{\rm{c}}2}}}\Psi_{\rm{sym}}({\bm r}) , \hspace{5mm}\tilde{\bm A}({\bm r})={\bf{0}}
\label{ini}
\end{align}
into  the Riccati form of Eq.\ (\ref{Ei1}),
where $\Delta_{T}$ is the energy gap at $\bar B=0$, $B_{{\rm{c}}2}=\mu_0 H_{{\rm c}2}$ is the upper critical field 
obtained by applying Helfand-Werthamer theory \cite{HW,KA04} to the present cylindrical Fermi surface, 
and $\Psi_{\rm{sym}}$ is Abrikosov's solution of the linearized Ginzburg--Landau equations in a symmetric gauge without prefactors.\cite{KitaText,KitaGL}
We then apply the method of solving Eq.\ (\ref{Ei1}) for an isolated vortex\cite{KitaText} to a circle of radius $R$  ($\gg a_2$) with many unit cells,
 focusing our attention to a single unit cell in the central region.
To be more specific, we start the integration from $f=0$ on the boundary with the hope that it will produce the relevant periodic solution 
in a unit cell far from the boundary because of the periodicity of the source term in Eq.\ (\ref{ini}).
The resulting $f$ in the unit cell is subsequently used to update $\Delta$ and $\tilde{\bm A}$ by Eqs.\ (\ref{Ei3}) and (\ref{Ei4}), which are then connected outside by Eq.\ (\ref{sym}) and substituted into Eq.\ (\ref{Ei1}) for the next iteration. 
The convergence of the iteration can be checked by monitoring the free energy\cite{Eilenberger,Klein,Ichioka2} of the unit cell.
We confirmed that the free energy decreases as the iteration proceeds,
which was stopped when the relative difference between the old and new free energies decreased to below $5.0\times10^{-4}$.
We also checked that choosing $R\gtrsim 4a_2$ brings about excellent convergence with respect to $R$ .

Functions $g$ and ${\bm B}={\bm\nabla}\times {\bm A}$ thereby obtained are used 
to construct the source term on the right-hand side of Eq.\ (\ref{el}).
The differentiation with respect to ${\bm p}_{\rm F}$ can be performed numerically without difficulties for the present cylindrical Fermi surface.
Equation (\ref{el}) with the source term is then solved by the finite-difference method\cite{Press07} to obtain ${\bm E}({\bm r})$ in the vortex lattice,
which is substituted into Gauss' law $\rho({\bm r})=\epsilon_0{\bm\nabla}\cdot{\bm E}$ to find the charge distribution.
We checked that the charge neutrality within the unit cell is satisfied.

Our results below were obtained for  $\delta\equiv 1/k_{\rm F}\xi_0=0.001$, $\lambda_{\rm{TF}}/\xi_0=0.001$,
and $\lambda_{\rm L}/\xi_0=5.0$ unless stated otherwise,
where $\xi_0\equiv \hbar v_{\rm F}/\Delta_0$ and $\lambda_{\rm{L}}\equiv\sqrt{\hbar/\mu_0\Delta_0\xi_0e^2 N(0) v_{\rm{F}}}$
are the coherence length and London penetration depth, respectively, defined in terms of the zero-temperature energy gap $\Delta_0$ at $H=0$.
The temperature and charge density were normalized by the superconducting transition temperature $T_{\rm{c}}$  at $H=0$
and $\rho_0\equiv\Delta_0\epsilon_0/|e|\xi^2_0$, respectively.

\begin{figure}[b]
        \begin{center}
                \includegraphics[width=0.9\linewidth]{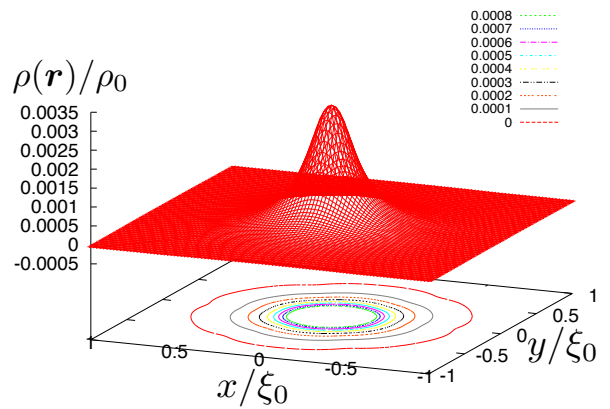}
                \end{center}
                                \caption{(Color online) Charge density $\rho({\bm r})/\rho_0$ in the core region $-1\leq x/\xi_0\leq1$ for $\bar{B}/B_{{\rm{c}}2}=0.073$ at $T/T_{\rm{c}}=0.2$,
                                where  $a_2=3.81\xi_0$.}
\label{fig1}
\end{figure}
\begin{figure}[b]
        \begin{center}
                \includegraphics[width=0.9\linewidth]{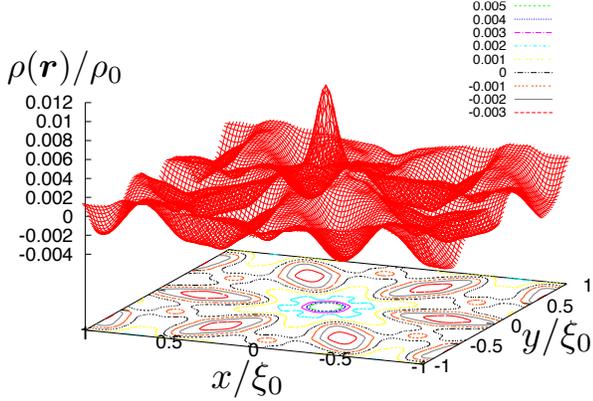}
                \end{center}
\caption{(Color online) Charge density $\rho({\bm r})/\rho_0$ in the core region $-1\leq x/\xi_0\leq1$ for $\bar{B}/B_{{\rm{c}}2}=0.51$ at $T/T_{\rm{c}}=0.2$, where  $a_2=1.44\xi_0$.}
\label{fig2}
\end{figure}

Figures \ref{fig1} and \ref{fig2} show spatial variations of the charge density  around a vortex core at $T/T_{\rm{c}}=0.2$ calculated for $\bar{B}/B_{{\rm{c}}2}=0.073$ and $0.51$, respectively.  The charge in Fig.\ \ref{fig1} is distributed isotropically with a conspicuous peak at the core center, indicating that this vortex in a low magnetic field is almost isolated. 
In contrast, the distribution at $\bar{B}/B_{{\rm{c}}2} =0.51$ strongly reflects the triangular symmetry of densely packed vortices.
Moreover, we observe that the peak height at the core center is enhanced by an order of magnitude from the case of $\bar{B}/B_{{\rm{c}}2}=0.073$.

\begin{figure}[b]
        \begin{center}
                \includegraphics[width=0.9\linewidth]{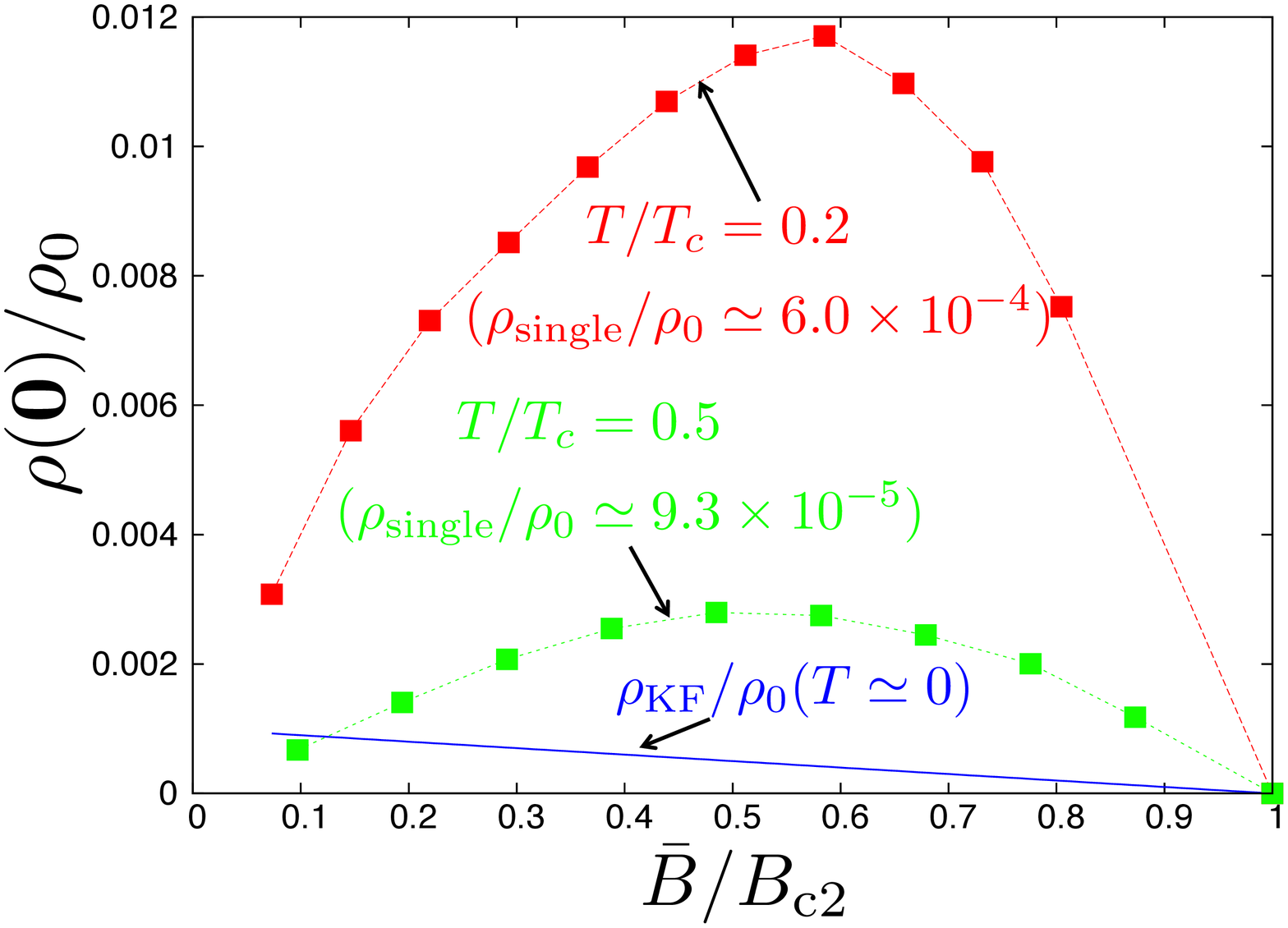}
                \end{center}
                \caption{(Color online) Normalized charge density at the vortex center,  $\rho({\bf{0}})/\rho_0$, as a function of magnetic field calculated for $T/T_{\rm{c}}=0.2$ (red line) and $T/T_{\rm{c}}=0.5$ (green line). 
                The field dependence of $\rho_{\rm{KF}}/\rho_0$ obtained from Eq.\ (\ref{Khomskii}) (blue line) is also plotted.}
      \label{fig3}
      \end{figure}

Figure \ref{fig3} shows the field dependence of the charge density $\rho({\bf{0}})$ at the vortex center for $T/T_{\rm{c}}=0.2$ and $0.5$.
In both cases, $\rho({\bf{0}})$ initially increases as the magnetic field is increased, reaches its maximum around $B_{\rm{c}2}/2$,
and decreases thereafter towards zero at $B_{{{\rm c}}2}$. 
The initial increase can be attributed to the factor ${\bm B}$ in Eq.\ (\ref{el}), whereas the latter decrease is due to the factor $\langle \partial g/\partial {\bm p}_{\rm F}\rangle_{\rm F}$, which decreases with decreasing pair potential.
We found that the maximum value of $\rho({\bf{0}})$ near $B_{{{\rm c}}2}/2$ can be about $10$--$10^2$ times larger than
the peak value $\rho_{\rm single}$ at $H_{{\rm c}1}$ obtained by solving Eqs. (\ref{Ei}) and (\ref{el}) for an isolated vortex.
Thus, vortex charging due to the Lorentz force has a strong field dependence with a peak and can be enhanced substantially
from the value for an isolated vortex. This is the main finding of the present study.

This field dependence of the vortex-core charge can also be reproduced analytically near $T_{\rm c}$. 
In this region, it is possible to expand the quasiclassical Green's functions $(f,g)$ in $\Delta$ 
by regarding the gradient operator as $O(|\Delta|^1)$ and retaining terms up to $O(|\Delta|^3)$.\cite{KitaText} 
Equation (\ref{el}) is thereby transformed into
\begin{equation}
(-\lambda^2_{\rm{TF}}\bm{\nabla}^2+1){{\bm E}}={\bm B}\times \underline{R}_{{\rm{H}}}{\bm j},\label{elGL}
\end{equation}
where 
\begin{equation}
\underline{R}_{{\rm{H}}}\equiv\frac{1}{2eN(0)}\left\langle\frac{\partial \bm{v}_{\rm{F}}}{\partial {\bm{p}}_{\rm{F}}}\right\rangle_{\rm{F}}\langle\bm{v}_{\rm{F}}\bm{v}_{\rm{F}}\rangle^{-1}_{\rm{F}}\label{Hall}
\end{equation}
is the normal Hall coefficient.
In particular, for a two-dimensional system with an isotropic Fermi surface,  Eq. (\ref{elGL}) can be rearranged into
\begin{equation}
(-\lambda^2_{\rm{TF}}\bm{\nabla}^2+1){{\bm E}}=\frac{1}{2eN(0)\varepsilon_{\rm{F}}}{\bm B}\times{\bm j},\label{elGL1}
\end{equation}
where $\varepsilon_{\rm{F}}$ is the Fermi energy of a two-dimensional free-electron model. 
We then substitute the expression for ${\bm j}$ near $T_{\rm c}$ \cite{KitaText} into Eq.\ (\ref{elGL1}) 
and  assume the term with $\lambda_{\rm{TF}}$ $(\ll\xi_0)$ to be negligible near $T_{\rm c}$.
Subsequently, we insert the resulting expression for ${\bm E}$ into Gauss' law $\rho({\bm r})=\epsilon_0{\bm\nabla}\cdot{\bm E}$ 
and approximate ${\bm B}({\bm r})\approx\bar{\bm B}$ as appropriate near $H_{{\rm c}2}$.
We thereby obtain
\begin{equation}
\rho({\bm r})\simeq\frac{7\zeta(3)\hbar v_{\rm{F}}^2\epsilon_{0}}{32(\pi k_{\rm{B}} T)^2 \varepsilon_{\rm{F}}} \bar{B}\bm{\nabla}^2|\Delta|^2,\label{elGL2}
\end{equation}
where $\zeta(3)=1.202\cdots$ is the Riemann zeta function. 
Now, the pair potential around the vortex core located at the origin may be approximated as 
\begin{subequations}
\label{Delta-core}
\begin{align}
\Delta\simeq\Delta_{\rm{max}}\sqrt{x^2+y^2}/\xi_{\rm{c}} ,
\label{Delta-core1}
\end{align}
where $\xi_{\rm{c}}$ is the vortex core size and $\Delta_{\rm{max}}$ denotes the maximum value of the pair potential in the vortex lattice.
We may also express $\Delta_{\rm max}$ on the basis of Abrikosov's theory\cite{KitaText,Ab} as 
\begin{align}
\Delta_{\rm{max}}\simeq \alpha \Delta_0\sqrt{\frac{B_{{\rm{c}}2}-\bar{B}}{B_0}},
\label{Delta_max}
\end{align}
where
\begin{align}
\alpha\equiv\sqrt{\frac{3^{\frac{1}{4}}}{4\left(1-\frac{\xi_0^2}{4\lambda_{\rm{L}}^2}\beta\right)I_{00}^{(4)}}}, \hspace{10mm} B_0\equiv\frac{\hbar}{2|e|\xi_0^2}
\end{align}
\end{subequations}
with $\beta\equiv{7\zeta(3)\Delta_0^2}/{8(\pi k_{\rm{B}} T)^2}$ and  $I_{00}^{(4)}=1.16$.
Substituting Eq.\ (\ref{Delta-core}) into Eq.\ (\ref{elGL2}), we obtain
\begin{equation}
\rho({\bf{0}})\simeq\frac{\rho_0}{B_0^2}\frac{\alpha^2\beta\delta}{({\xi_{\rm{c}}}/{\xi_0})^2} \bar{B}(B_{{\rm{c}}2}-\bar{B}).
\label{elGL4}
\end{equation}
Equation (\ref{elGL4}) with Eq.\ (\ref{Delta_max}) indicates that the magnetic field dependence of $\rho({\bf{0}})$ is determined by the competition between the increasing magnetic field and the decreasing  pair potential as $\rho({\bf{0}})\propto\bar{B}\Delta^2_{\rm{max}}$.
Although it was derived near $T_{\rm c}$,  Eq.\ (\ref{elGL4}) reproduces the key features of the field dependence of the core charge at low temperatures given in Fig.\ \ref{fig3}.

As already mentioned, alternative mechanisms for the core charging have been proposed by Khomskii and Freimuth\cite{KF} and Feigel'man and coworkers\cite{Feigel'man,Blatter} 
based on considerations of an isolated vortex.
Although different from each other in detail, the two mechanisms have a common feature that the reduction of $|\Delta({\bf r})|$ in the presence of a finite slope in the normal density of states is the key source for the vortex-core charging; 
they have little connection with the Lorentz force.
Hence, charge accumulation in the core region by these mechanisms should decrease monotonically as $H$ is increased.
For example, the core charge density given by
Khomskii and Freimuth\cite{KF}  can also be expressed as
\begin{equation}
\rho_{\rm{KF}}/\rho_0\simeq {\delta\Delta^2_{\rm{max}}}/{\Delta^2_0}\sim O(\delta).\label{Khomskii}
\end{equation}
We also plot $\rho_{\rm{KF}}/\rho_0$ in Fig.\ \ref{fig3}, which is seen to decrease monotonically in marked contrast with the peak structure of $\rho({\bf 0})/\rho_0$.
In other words, the Lorentz force mechanism for vortex-core charging can be experimentally distinguished from the other two mechanisms in an unambiguous manner
by observing the magnetic-field dependence.
Note also that the peak value of the charge density $\rho({\bf 0})$ is about 10 times larger than  $\rho_{\rm{KF}}$ at $T/T_{\rm{c}}=0.2$.

\begin{figure}[b]
        \begin{center}
                \includegraphics[width=0.9\linewidth]{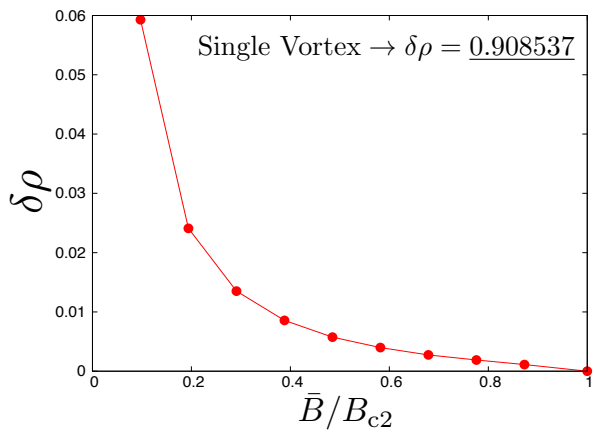}
                \end{center}
\caption{
(Color online) Relative difference  in the charge density at the vortex center between $\lambda_{\rm{L}}=5\xi_0$  $[\equiv\rho({\bf{0}})_5]$ and $\lambda_{\rm{L}}=20\xi_0$ $[\equiv\rho({\bf{0}})_{20}]$,
where $\delta\rho$ is defined as $[\rho({\bf{0}})_5-\rho({\bf{0}})_{20}]/\rho({\bf{0}})_5$.
}
\label{fig4}
\end{figure}

Next, we turn our attention to the $\lambda_{\rm{L}}$ dependence of vortex charging. Figure \ref{fig4} plots the relative difference in the charge density at the vortex center
between $\lambda_{\rm{L}}=5\xi_0$ and $\lambda_{\rm{L}}=20\xi_0$ at $T/T_{\rm{c}}=0.5$, which we denote by $\delta\rho$, as a function of $\bar B/B_{{\rm c}2}$.
Thus,  $\delta\rho$ is of order $0.1$ even at low fields and decreases rapidly as the field is increased,
indicating that we can neglect the  $\lambda_{\rm{L}}$ dependence of the charge density to the first approximation
when considering it as a function of $\bar B/B_{{\rm c}2}$.
In this context, it should be noted that charge accumulation due to the Lorentz force in {\it an isolated vortex} has a substantial $\lambda_{\rm L}$ dependence;
indeed, $\delta\rho$ can be of order 1.
This is because $\lambda_{\rm L}$ strongly affects the magnitude of ${\bm B}({\bm r})$ around the core\cite{KitaText,Ab} and hence the core charge according to Eq.\ (\ref{el}).

Finally, we estimate the order of the accumulated charge around a vortex from Eq.\ (\ref{elGL4}).
Although derived near $T_{\rm c}$, we verified that the formula reproduces the numerical results in Fig.\ \ref{fig3} 
reasonably well over $T\gtrsim 0.2T_{\rm c}$. 
Let us approximate the charge profile near the core center in Fig.\  2 as a cone of radius  $0.5\xi_0$ with the peak value in the simulation. 
We thereby obtain a rough estimate for the peak value of the accumulated charge $Q_{\rm{unit}}$
in the core region of radius $0.5 \xi_0$ per length $\Delta z=10$ $\AA$ along the flux line 
for $T/T_{\rm{c}}\lesssim0.2$ as
\begin{align}
Q_{\rm{unit}}&\equiv\frac{\pi}{12}\Delta z \xi_0^2 \rho({\bf{0}})_{\rm{max}}
\approx\pi \Delta z \xi_0^2 \delta \rho_0
=\eta\delta\Delta_0|e| ,\label{estimate}
\end{align}
where $\eta\equiv\pi\epsilon_0\Delta z/|e|^2\simeq1.09\times10^{18}$ ${\rm{J^{-1}}}$. 
For the metallic superconductor Nb, we adopt  $k_{\rm{F}}\simeq1.0$ $\AA$,  $\Delta_0\simeq1.4$ ${\rm{meV}}$, and $\xi_0\simeq380$ $\AA$\cite{OKlein} for Eq.\ (\ref{estimate}) to obtain $Q_{\rm{unit}}\simeq 5 \times10^{-7}|e|$ C. 
For the high-$T_{\rm c}$ superconductor YBa$_2$Cu$_3$O$_{7-x}$, 
we may substitute $k_{\rm{F}}\simeq1.0$ $\AA$, $\Delta_0\simeq28$ meV,\cite{H.L} and $\xi_0\simeq30$ $\AA$\cite{Kumagai} into Eq.\ (\ref{estimate}), thereby obtaining $Q_{\rm{unit}}\simeq1\times10^{-4}|e|$ C.
As can be seen from Eq.\ (\ref{estimate}), $\delta$ and $\Delta_0$ are crucial elements determining the magnitude of the accumulated charge. 
In this context, the parameter $\delta$ of the iron-based layered superconductor FeSe has been reported to reach as high as  $\delta\sim1$.\cite{FeSe}
Hence, we now have a greater chance of clearly observing the field dependence.
Among the possible experimental methods to this end are nuclear magnetic resonance\cite{Kumagai,Mounce} and the Kelvin method for measuring
 the Hall voltage.\cite{Kel1,Kel2}
 Finally, it should be noted that the Fermi surface curvature is another crucial element determining the magnitude and sign of the vortex-core charge.\cite{Kita09,Ueki}

In summary,  we have clarified the magnetic field dependence of the vortex-core charge due to the Hall effect
using augmented quasiclassical equations of superconductivity both numerically and analytically. 
We found a peak structure characteristic of the Lorentz force mechanism for charging
that should be common among all type-II superconductors.
Its observation to confirm the Lorentz force on persistent currents is a challenging topic.

\end{document}